\documentclass{article}
\usepackage{arxiv}

\usepackage[utf8]{inputenc} 
\usepackage[T1]{fontenc}    
\usepackage{hyperref}       
\usepackage{url}            
\usepackage{booktabs}       
\usepackage{amsfonts}       
\usepackage{nicefrac}       
\usepackage{microtype}      
\usepackage{floatrow}
\usepackage{doi}
\usepackage{graphicx}
\usepackage{subcaption}

\begin{document}

\title{Determining the origin of impulsive noise events using paired wireless sound sensors}

\author{
	Fabian Nemazi \\
	Norwegian University of Life Sciences \\
	\And
    Jon Nordby \\
	Soundsensing AS \\
	\texttt{jon@soundsensing.no} \\	
}

\date{}

\maketitle
\renewcommand{\abstractname}{\vspace{-\baselineskip}} 

\begin{abstract}	\noindent
This work investigates how to identify the source of impulsive noise events using a pair of wireless noise sensors.
One sensor is placed at a known noise source, and another sensor is placed at the noise receiver.
Machine learning models receive data from the two sensors and estimate whether a given noise event originates from the known noise source or another source.
To avoid privacy issues, the approach uses on-edge preprocessing that converts the sound into privacy compatible spectrograms.
The system was evaluated at a shooting range and explosives training facility, using data collected during noise emission testing.
The combination of convolutional neural networks with cross-correlation achieved the best results.
We created multiple alternative models using different spectrogram representations.
The best model detected 70.8\% of the impulsive noise events and correctly predicted 90.3\% of the noise events in the optimal trade-off between recall and precision.\\


\end{abstract}

\section{Introduction}


Environmental noise pollution is a major issue in many urban environments.
Some types of noise sources, such as industry, construction, shipyards and shooting ranges, originates from a particular location, and is the responsibility of the site operator.
To investigate what influence a noise source has on potentially affected neighbors,
one must measure the noise levels at the receiver,
while isolating out only the noise that originates from the source under investigation.
When the receiver is in an urban area, there also tends to be many other time-varying sources present, such as car traffic, airplane traffic, other construction noise and noise from social activities.
This can make it challenging to determine whether a noise event observed at a receiver originated from the source under study, or some unrelated noise source.
Often this requires considerable amount of manual inspection and skilled data analysis,
which is very costly for monitoring over longer periods of time.

To tackle this challenge, we propose an automated system consisting of two wireless acoustics sensors connected to a server, and a machine learning model to analyze the data.
One sensor is positioned at the noise source under investigation, and the other at the receiver location of interest.
The machine learning model analyzes the data from the two sensors, and determines whether a noise event observed at the receiver originated at the source of interest.

This solution has been tested in a case-study, using data captured at a shooting range and explosives training facility.

\noindent

\newpage
\section{Background}

\subsection{Politiets Nasjonale Beredskapssenter}

Politiets Nasjonale Beredskapssenter (PNB) is a combined training facility and operative base for the police special forces in Norway.
The site consists of an office building, an airport for police helicopters
and training facilities with indoor and outdoor shooting ranges for rifles and handguns, as well as buildings for practicing breach and entry with explosives.

A key criterion for the location of the center was proximity to Oslo city and urban area,
in order to ensure fast response times.
Another criterion was a location far from residential areas, to prevent disturbance caused by noise from shooting and explosives used during training.
The chosen location combines these two requirements:
It is within 20 minutes drive to downtown Oslo, and noise is managed by extensive noise abatement measures\cite{RieberPNBForbedredeTiltak}.
Some examples of the constructed measures:
Training areas are surrounded by 10 meter tall berms, with additional 2 meter tall
noise dampening fences.
For the outdoor shooting ranges, the shooting positions are inside a building that is closed on all sides apart from the shooting direction.
Buildings used for explosives training are fitted with noise absorbing walls and interiors.

\subsection{Regulations}

In Norway the recommended noise limits for shooting-ranges (and other kinds of environmental noise) is defined in T-1442/2016~\cite{T-1442/2016}.
It defines two zones, a red zone which is not suitable for noise sensitive buildings,
and a yellow zone where noise sensitive buildings may be placed only if measures are made to provide acceptable noise conditions.  
For shooting-ranges the red zone is defined by $L_{den}~$\cite{EuNoiseDirective} of 35 dB and $L_{AFmax}~$\cite{IECSoundLevelMeters} of 65 dB,
and the yellow zone is defined by $L_{den}$ 45 dB and $L_{AFmax}$ 75 dB.
At night (between 23:00 and 07:00) no activity should be allowed.

The local municipality use these recommendations as a starting point,
and define the local regulations though a zoning plan~\cite{ZoningPlanNorway}.
In the case of PNB, concern for noise from the local residents was very high.
For this reason the authorities set much stricter regulations~\cite{pnb-reguleringsplan} regarding noise from training activities than the recommended noise limits found in T-1442/2016~\cite{T-1442/2016}.
The additional restrictions include:

\begin{enumerate}
  \item Training activity is only permitted on Monday - Friday, between 07:00 and 19:00.
  \item The number of explosive charges outdoors may not exceed 1250 per year.
  \item Maximum number of gunshots per year of 2 000 0000.
  \item The yellow and red zone may not exceed the original approved plans.
\end{enumerate}

\subsection{Solution requirements}

The continuous monitoring of noise requires the use of noise sensors
that may operate outside for long periods of time.
These should regularly report the necessary data to a central storage server,
so that the data can be accessed by site operators.
For privacy reasons, audio or other speech information should not be transmitted from (or stored on) the wireless sensors.
For this reason, the sensor should process the audio data on the sensor, and only send privacy-compatible acoustical information to the cloud.
Earlier works have shown that spectrograms with an integration time of at least 125 milliseconds does not have intelligible speech \cite{gontier_efficient_2017:censecoder}.
The Soundsensing dB20 sensor and data platform fulfill these requirements.

When using two sensors that operating independently,
their clocks may drift somewhat with respect to each other.
This means that the data streams across devices will not be perfectly aligned,
and an analysis should accommodate this.
In this work, we allow for one sound sensor to be up to $\pm 10$ seconds behind/ahead of the other.

\subsection{Related work}

The task of detecting short sound events (\emph{Sound Event Detection}) has featured in several challenges as part of the Detection and Classification of Audio Scenes and Events (DCASE) workshop series \cite{stowell_detection_2015}\cite{lafay_sound_2017} \cite{mesaros_sound_2019}. 
Gunshots are a part of several datasets for Environmental Sound Classification such as Urbansound8k \cite{salamon_dataset_2014:urbansound8k} and ESC-50 \cite{piczak2015dataset:esc50},
and many machine learning models have been investigated on these datasets.
Gunshot detection has also been studied in the context of security systems \cite{valenzise_scream_2007}.

An earlier work by the authors, using the same site (PNB) and dataset, focused on the detection of noise events at the source location \cite{jonnordby2021}.
The results presented here are also covered in a master thesis by the primary author\cite{fabiannemazi}.


\section{Methods}

\subsection{Data collection}
The data was collected while testing the noise abatement measures,
when construction of the site was nearly completed.
In such tests, each tested configuration has the weapons fired every thirty seconds, using five or ten repetitions \cite{Veileder-M128/2014}.
The data collection was done on four different days,
two days focusing on noise propagating to the north,
and two days on noise propagating to the west.
The receiver locations and weather conditions were such that sound propagation to the measured area was maximized.
Many weapon configurations were tested at each of the 3 different shooting ranges,
and 2 locations for explosives training.
There was an emphasis on the weapon configurations and locations that create the most noise.
This makes the data representative of worst-case conditions in terms of single-event noise experienced by neighbors.

\begin{figure}[htb!]
    \centering
    \includegraphics[width=14cm]{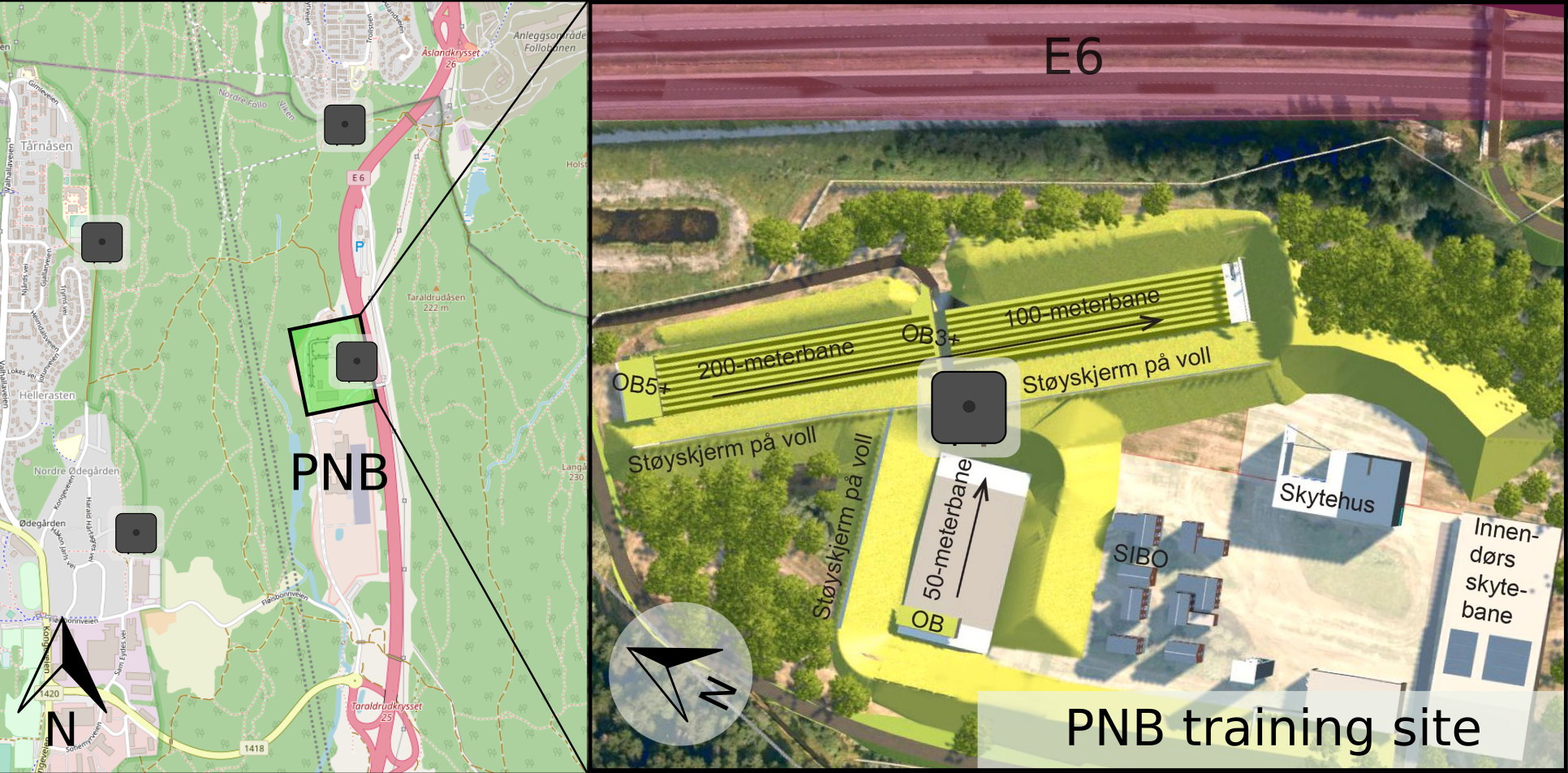}
    \caption{Device locations used in model development. Left: Overview, with \textit{receiver} locations in residential areas to north and west of site. Right: The training facility and \textit{source} device location in the center.
    }
    \label{fig:data_collection_locations}
\end{figure}

Professional sound recorders were used for collecting the audio data.
Recorders were positioned at multiple locations at the training facility,
and in the surrounding area towards the residential area, as shown in Figure \ref{fig:data_collection_locations}.
Neighbors at the settlement were notified ahead of the tests and data collection,
and signs with information were in place during data collection.
The recorders were placed away from areas where pedestrians abide, to avoid recording of speech.

During data collection, the time and details of each test activity was noted down,
including the weapon configuration, firing location, orientation etc.

\subsection{Data labelling}

The recording unit at the center of the site was chosen as the reference data stream.
The labels for each individual noise event (start and end time) was initially generated
by using an unsupervised Hidden Markov Model \cite{rabiner1986:HMM} with two states model on the sound levels. 
These timestamps were then reviewed and adjusted based on manually listening to the audio. 

The recordings of all the other devices was then reviewed,
manually comparing the audio with the timestamps of the labeled events.
There were some offsets in time between the different devices,
and these were compensated for by shifting the data in time to align them with the reference device.

Noise events at the receiver was labeled with 3 levels of how easily the noise could be heard: not heard, faint and clear.
The labeling was done by listening to the recordings, helped by a spectrogram image and the knowledge that each event was spaced 30 seconds apart.
We note that this kind of labeling is highly subjective and contextual,
and that this kind of active, informed and critical listening is likely to overestimate how noticeable a noise event
would be compared to an everyday observer in the residential area.
For this reason, the labels as binary, considering clear as a noise-event, and faint and not heard as absence of a noise-event.

\subsection{Sound Event Detection on individual sensors}

\begin{figure}[htb!]
    \centering
    \includegraphics[width=14cm]{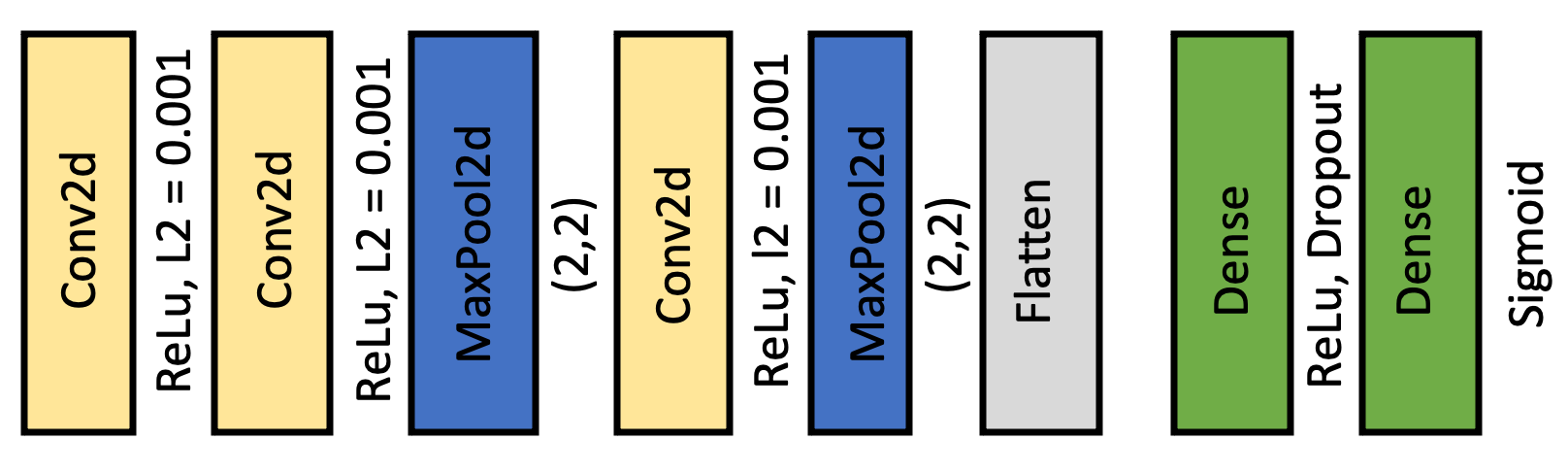}
    \caption{Architecture of CNN models used to detect noise events.}
    \label{fig:cnn_model}
\end{figure}

To detect a noise event in a single device data stream, a Convolutional Neural Network (CNN) was used.
The base architecture can be seen in figure \ref{fig:cnn_model}.
Two models were trained independently,
one with data from the noise source (center of shooting ranges),
and the other with data from the noise receiver (near residential area).

The CNN model receives as input a 1-second spectrogram window with 125 ms resolution.
Windows are generated with overlap, using a 0.375 seconds hop.
The spectrogram is computed using  Short-Time-Frequency-Transform (STFT) 
and processed with a triangular Mel-spaced filter bank of with 30 bins and then converted to decibels.
Windows that have at least one noise event is given a positive label, and other windows are given a negative label.
A few different spectrogram variations was tested as input to the CNN,
by adding the first, second and third order difference as channels (delta stack),
together with the base channel with the original mel-spectrogram (mel spec).

\subsection{Combining detected noise events to determine origin}

The outputs of the two Convolutional Neural Network models is a time-series of predictions range $0.0 - 1.0$, representing the probability of a noise event.
The streams for both devices are divided into overlapping windows with a fixed duration of 26 seconds, allowing for up to 13 seconds of time misalignment (from clock drift).
\newline

\begin{figure}[htb!]
    \centering
    \includegraphics[width=14cm, height=8cm]{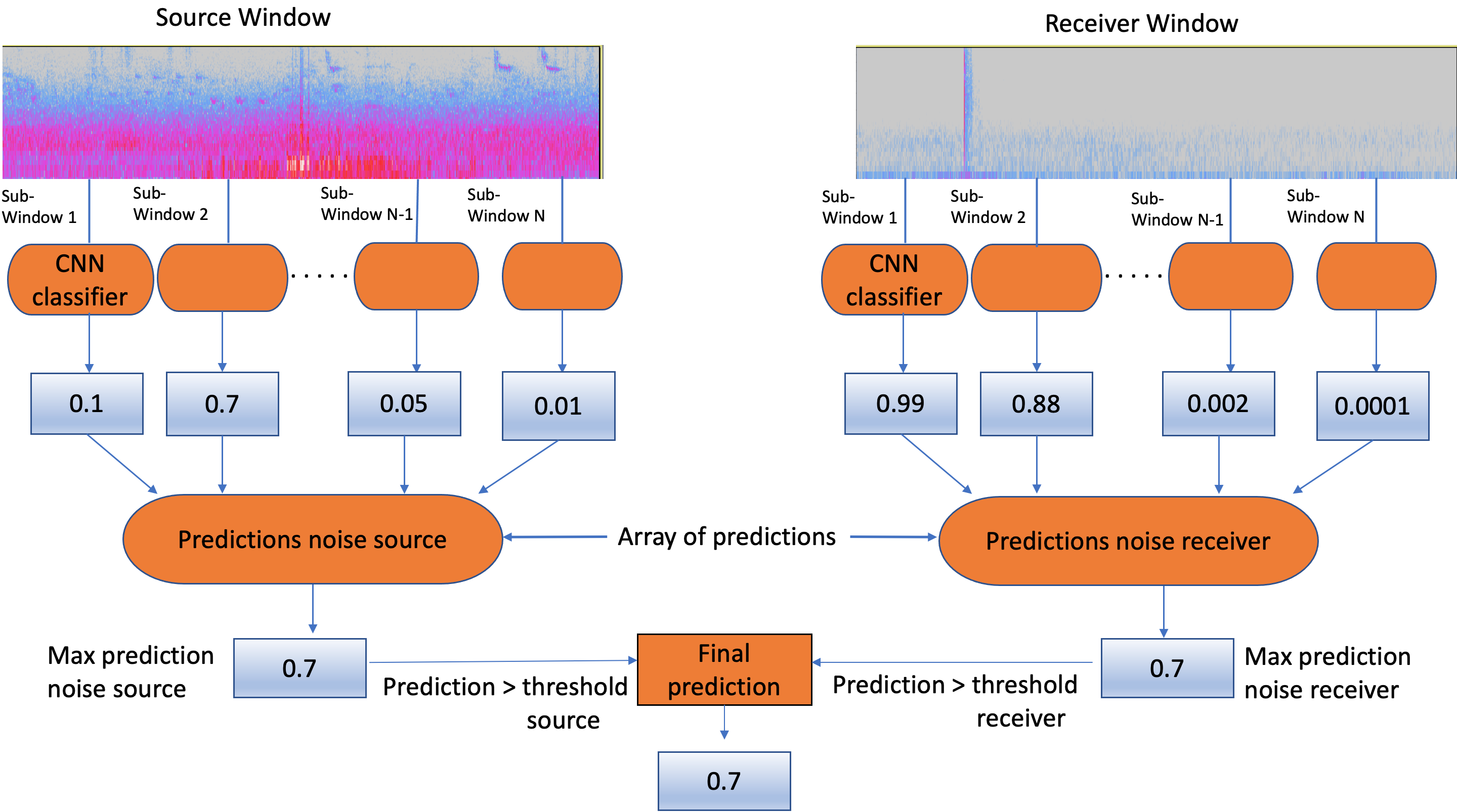}
    \caption[Bundled model]{Combining two convolutional neural networks to predict noise origin. Each model gives a time-series of predictions for each 26 seconds window.
    The highest probability of the predictions is extracted.
    Max aggregation is also used between the value from each of the two series.
    }
    \label{fig:model_architecture}
\end{figure}

Figure \ref{fig:model_architecture} illustrates the process of combining the two CNN models.
The steps could be summarized in the following way:
\begin{enumerate}
    \item If the noise source model detects no activity (noise) at the training facility, noise around the noise receiver must have another origin. The final prediction for the window is 0.
    \item If the noise source model detects activity (prediction $>$ \emph{threshold source}), the maximum prediction from the noise receiver model is collected.
    \item If the noise receiver model has a maximum prediction higher than \emph{threshold receiver}, it is considered the representative prediction for the window.
    \item If the maximum prediction is lower than \emph{threshold receiver}, calculate the maximum cross-correlation between the predictions from both models. The maximum cross-correlation between the prediction stands as the final prediction for the window.
\end{enumerate}


Two variations of the combined model was created.
These variations are denoted \emph{bundled one} and \emph{bundled two}.
The \emph{bundled one} model was a combination of the noise receiver model with mel-spectrogram representation and the noise source model trained on 3 delta stack-spectrogram representation.
The \emph{bundled two} model combined a noise receiver model with 1 delta stack-spectrogram representation and noise source model with 3 delta stack-spectrogram. 

Thresholds were set based on performance on the validation set, with the best \emph{threshold receiver} being 0.6 and \emph{threshold source} being 0.8.

\newpage
\section{Results}

Performance of the individual noise event detection models that were tested can be found in tables \ref{tab:source_results} and \ref{tab:receiver_results}.
At the source, all model variations achieve good detection performance.
The 3 Delta stack spectrogram representation achieves the best average precision (AP) score with 90.1\%.

At the receiver, the detection performance is considerably lower.
1 Delta stack achieves the best score (69.3\% AP) and is followed by the mel-spec model.


\begin{table*}[!htb]

  \begin{floatrow}[2]
     
     \begin{subtable}{0.3\linewidth}
     \begin{tabular}{lr}
        \hline\hline
        Model & AP  \\ 
        \hline
        Mel-spec & 89.2  $\pm$ 1.1 \% \\ 
        1 Delta stack & 89.3 $\pm$ 1.1 \% \\
        2 Delta stack & 88.9 $\pm$ 1.6 \% \\
        3 Delta stack & \textbf{90.1 $\pm$ 2.3 \% } \\
        \hline
    \end{tabular}
    \caption{results for source model}
    \label{tab:source_results}
    \end{subtable}
    
    \hspace{3cm}%

    \begin{subtable}{0.3\linewidth}
    \begin{tabular}{lr}
        \hline\hline
        Model & AP  \\  
        \hline
        Mel-spec & 66.5  $\pm$ 27.8 \% \\ 
        1 Delta stack & \textbf{69.3 $\pm$ 23.3 \%} \\
        2 Delta stack & 63.4 $\pm$ 27.3 \% \\
        3 Delta stack & 62.0 $\pm$ 28.3 \% \\
        \hline
    \end{tabular}
    \caption{results at receiver}
    \label{tab:receiver_results}
    \end{subtable}

  \end{floatrow}
  
  \label{tab:individual_results}
  \caption{Performance of CNN models at source and at receiver locations. Best models marked in bold.}
\end{table*}

\begin{figure}[htb!]
    \centering
    \includegraphics[width=10cm, height=7.5cm]{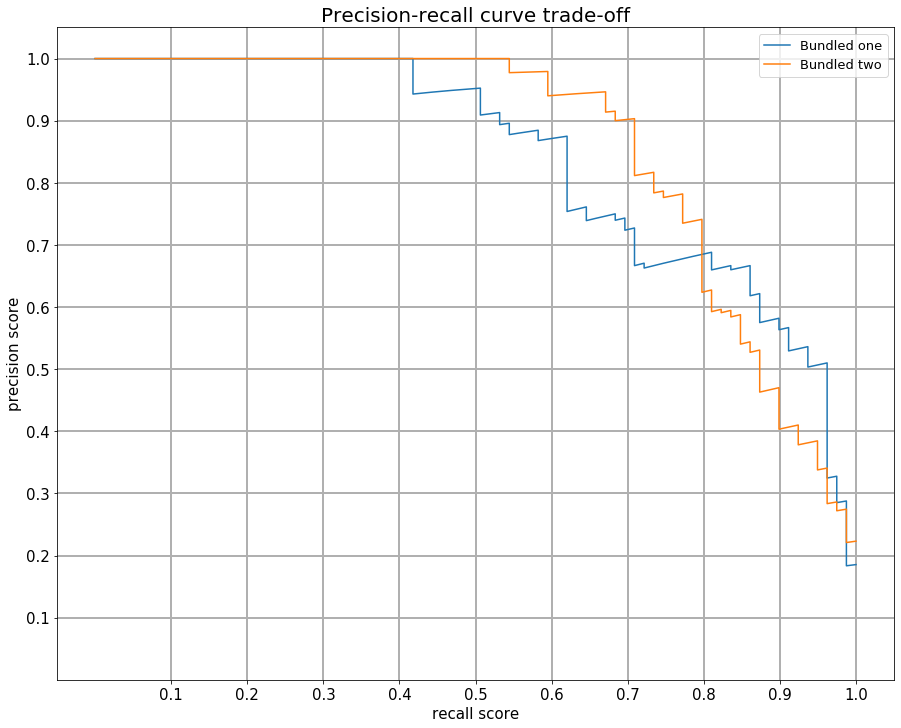}
    \caption[Precision-recall curve for the bundled models on test data]{Precision-recall curve for the bundled models. The labels \emph{bundled one} and \emph{bundled two} represent the bundled solutions.}
    \label{fig:nn_cross_test_scores}
\end{figure}

The performance for the combined model scores are listed in table \ref{tab:combined_results}.
The precision/recall trade off is visualized in figure \ref{fig:nn_cross_test_scores}.

\renewcommand{\arraystretch}{1}
\begin{table}[htb!]
    \centering 
    \scalebox{1}{
        \begin{tabular}{lrrr}
            \hline\hline
            Model  & Precision & Recall & Optimal F1 \\
            \hline
            Bundled one & 90.3 \%    & 70.80 \% & 79.43 \% \\ 
            Bundled two & 66.6 \%    & 86.07 \% & 75.13 \% \\
            \hline
        \end{tabular}
    }
\caption{\label{tab:combined_results}The performance of the combined model. Evaluated on the test set.}
\end{table}

\section{Discussion}

\subsection{Individual CNN model performance}

At the noise source, the CNN model can easily detect the noise events, reaching AP above 90\%.
At the noise receiver, the performance is considerably lower, with AP between 62\% to 69\%.
This is largely expected, as the signal-to-noise ratio is significantly higher at the source than at the receiver.
In addition, the much shorter distance means lower variability due to variations from the atmosphere during sound propagation.
This means that the performance of the model at the noise-receiver is the bottleneck in terms of overall performance.

\subsection{Precision/recall tradeoff}

Figure \ref{fig:nn_cross_test_scores} showed that the \emph{bundled two} model achieved an F1 score of 79.43\%, followed by the \emph{bundled one} model achieving 75.13\%. The best model (\emph{bundled two}), which included 1 delta stacking in the settlement, achieved a better optimal trade-off by approximately 4\%.
The two combined models have precision/recall curves that are considerably different.
The \emph{bundled one} model detects fewer cases of impulsive noise events. However, it detects the events with higher precision.
The \emph{bundled two} model detects many cases of impulsive noise, but detects these with lower precision.

A goal in the project was to be able to detect events to be within $\pm$ 10\% of the real number, and these model models fall short of that, with false negatives/positives in the 25\% range.
Therefore, the solution was not yet good enough to act as a fully automated solution.
However, by tuning the models to a recall of $<$90\% and allowing false negatives of about 50\%, one can use the detection as a first stage in a data analysis pipeline.
Each event must then be verified by a human afterwards to
reach the required precision.

\subsection{Dataset limitations}
The present models are trained and evaluated on data
from acoustics tests, not ordinary training activity.
During training, it is expected that multiple people will use the shooting ranges concurrently, which will result in more events per time period, and occasionally multiple noise events at the same time.
This will cause more complex sound patterns, which may be harder to detect.
It is also expected that under normal conditions, noise will not spread as much, because the typical weapon configuration and weather conditions are not as noisy as during the worst-case testing.
This is good for neighbors, but may be harder for our system to detect.

\subsection{Further work}

In the months following this research,
sensors have been installed on location and the training activity has started.
The sensors provide continuous monitoring with automatic detection of noisy activity at the source, which can be tracked in a logbook.
In a planned follow-up study, sensors will also be placed into the surrounding residential area over several days, to capture any noise events that propagate there.
This updated dataset will allow the models proposed here to be re-trained
and evaluated under the ordinary training conditions.
The model will be used to semi-automatically find candidates for noise events originating at the source, which is then be manually verified.

In the current work, two sound event detection models were trained independently, and then a combined in a separate model.
A promising alternative would be to train the event detection sub-models jointly with the combined model, for example using a Siamese Network \cite{chicco2021siamese}.

\section{Conclusions}

In this work, a solution for automatically determining the origin of impulsive noise events was proposed, using two wireless sound sensors and a neural network to analyze the two data streams.
The presented solution handles privacy by pre-processing the data on-sensor into a privacy compatible spectrogram representation.
The proposed model was tested on data captured from a shooting range and training facility.
It was possible to achieve a high recall score at a precision that
is sufficient to be used for semi-automated analysis,
where detected events are human-verified afterwards.
However, the precision score at the desired recall is not judged sufficient for a fully-automated detection for continuous monitoring,
but this may be solvable with more work on datasets and model architecture.

\section*{Acknowledgements}
\noindent
Most of the work reported was done as part of a master thesis at Norwegian University of Life Sciences \cite{fabiannemazi}.

The data collection for this research was done in a collaboration project funded by Politiets Nasjonale Beredskapssenter construction project and Soundsensing AS.
In the first months of the project, Soundsensing received funding from the Research Council of Norway as part of the program STUD-ENT.

Organization of acoustical tests during the data collection was led by Lars R. Nordin at Brekke \& Strand AS.
Data collection for Soundsensing AS was performed by Erik Sjølund and Ole Johan Aspestrand Bjerke.
We thank the entire team for excellent facilitation and execution.

\newpage
\bibliographystyle{unsrt}
\bibliography{references.bib} 

\end{document}